\documentclass[11pt,twoside]{article}
\usepackage{asp2010}

\begin{document}

\title{The Challenge of Data Reduction for Multiple Instruments on Stratospheric Observatory For Infrared Astronomy (SOFIA)}
\author{Charcos-Llorens M. V.$^1$; Krzaczek R.$^2$; Shuping R. Y.$^3$; and Lin L.$^1$
\affil{$^1$Universities Space Research Association, NASA Ames Research Center, Moffett Field, CA 94035, USA}
\affil{$^2$Chester F. Carlson Center for Imaging Science, RIT, 54 Lomb Memorial Drive,  Rochester NY 14623, USA}
\affil{$^3$Space Science Institute, 4750 Walnut Street, Boulder, Colorado 80301, USA}}

\begin{abstract}
SOFIA, the Stratospheric Observatory For Infrared Astronomy, presents a number of interesting challenges for the development of a data reduction environment which, at its initial phase, will have to incorporate pipelines from seven different instruments developed by organizations around the world. Therefore, the SOFIA data reduction software must run code which has been developed in a variety of dissimilar environments, e.g., IDL, Python, Java, C++. Moreover, we anticipate this diversity will only increase in future generations of instrumentation. We investigated three distinctly different situations for performing pipelined data reduction in SOFIA: (1) automated data reduction after data archival at the end of a mission, (2) re-pipelining of science data with updated calibrations or optimum parameters, and (3) the interactive user-driven local execution and analysis of data reduction by an investigator. These different modes would traditionally result in very different software implementations of algorithms used by each instrument team, in effect tripling the amount of data reduction software that would need to be maintained by SOFIA.

We present here a unique approach for enfolding all the instrument-specific data reduction software in the observatory framework and verifies the needs for all three reduction scenarios as well as the standard visualization tools. The SOFIA data reduction structure would host the different algorithms and techniques that the instrument teams develop in their own programming language and operating system. Ideally, duplication of software is minimized across the system because instrument teams can draw on software solutions and techniques previously delivered to SOFIA by other instruments. With this approach, we minimize the effort for analyzing and developing new software reduction pipelines for future generation instruments. We also explore the potential benefits of this approach in the portability of the software to an ever-broadening science audience, as well as its ability to ease the use of distributed processing for data reduction pipelines.
\end{abstract}

\section{Introduction}
SOFIA is an airborne observatory designed primarily to carry out observations at infrared and sub-millimeter wavelengths that cannot be carried out from ground-based facilities.  SOFIA will host a variety of instruments observing in wavelength ranges from 0.3 to 600 microns which will be upgraded over time. This will produce a large diversity of data types which will likely increase as new generations of instruments are operated. 

The SOFIA Data Cycle System (DCS) \footnote[1]{see http://dcs.sofia.usra.edu} is a collection of tools and services that support both the General Investigator (GI) and the Science and Mission Operations staff from observation and mission planning, through observation execution on-board the aircraft, to data archiving and processing post-flight and distribution to the GI and the scientific community.  The DCS  will provide a uniform, extensible and supportable framework for all aspects of this data cycle. 

The DCS will support  data processing for both facility and Principal Investigator-class instruments, including archiving and pipelining of raw (Level 1), processed (Level 2), flux calibrated (Level 3), and higher level data products (e.g. mosaics and source catalogs). 
Data processing includes all steps required to obtain good quality flux calibrated data for spectroscopy, imaging, fast-acquisition, polarimetry, etc. Processing each data type requires a sequence of unique or common algorithms  with specific parameters to be tuned.  The DCS will incorporate, improve and maintain these algorithms which are provided by the instrument teams and  developed in a variety of  environments. In addition, these algorithms may require user-interaction or fine tuning of input parameters in order to return good quality data

\section{Concepts and Associations}
The DCS uses an Astronomical Observation Request (AOR) concept to collect up all needed information required to carry out an observation.  AORs are produced by the GI and SMO staff during the observation planning stage and then passed to the SI during flight for execution.  In addition, the AOR is the link between science and calibration data of the same observation type and defines the parameters necessary for post-processing. Therefore, it will identify the reduction pipeline and its parameters. For each level 2 product, the Pipeline Pedigree (PP) records the pipeline generating the data, the parameters, the processing date and the data involved in the process. AOR and PP concepts has been implemented and are operative in DCS. A similar concept will be necessary to track calibration activities. DCS will include a Flux Calibration Parameter (FCP) which will support the calibration process in order to document and reproduce the same results as needed. AOR, PP and FCP are characterized by unique key numbers that identify them as well as information about the data involved in the process.

\begin{figure}[htbp]
  \centering
    \includegraphics[width=5in, scale=0.5]{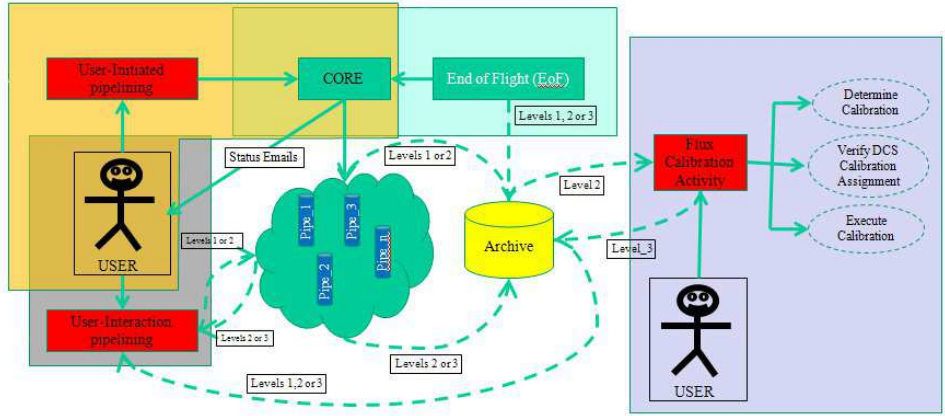}
    \caption{Data processing scenarios.} \label{arch}
\end{figure}

\section{Architecture}

The DCS will provide the framework for both automatic pipelining and human-in-the-loop processing. The DCS will host automatic pipelining at End-of-flight (EoF), user-initiated pipelining, and user-interactive processing and analysis.  Figure \ref{arch}
 illustrates how these scenarios relate to each other. The CORE is in charge of data processing within DCS (green actors). User can perform data processing outside DCS (red actors) and use DCS tools to extract and archive data. We show data fluxes as dashed arrows and process requests as plain arrows. We explain below the four main scenarios defining the data processing scenarios illustrated in Fig~\ref{arch}: 

\begin{itemize}
 \item {\bf EoF automatic pipelining producing immediate Level 2 products} [green area] \newline
 \small{Flight data, as for example raw observations or flight-processed products, are ingested at EoF. DCS calls pipelines automatically after ingestion of data observed during flight operation. Products from data reduction, typically level 2 data, are automatically archived as the data are processed, making them quickly available for scientific analysis.}

 \item {\bf Flux calibration producing Level 3 products} [purple area] \newline
 \small{Outstanding scientific results can be obtained only with flux calibrated data. Flux calibration is a complicated processes that is difficult to automate --- especially for an airborne observatory. The difficulty of defining a metric of the data quality makes necessary intervention of experienced scientists. Final Level 3 products can be archived in SOFIA database as well as their associated FCP.}
 
 \item {\bf User initiated processing and inspection (all levels)} [gray area] \newline
 \small{Pipelining can also be manually initiated by SMO scientists. For example, they may re-pipeline the data with modified parameters which would improve the quality of the final results, or when a new version of a pipeline becomes available.}
 
 \item {\bf User interactive processing (all levels)} [orange area] \newline
 \small{Likely, human intervention is often needed to verify results at any of the data levels. DCS will provide an interaction interface to extract data from the archive and run locally the same algorithms used during automatic pipelining. This allows the user to analyze the data at any step of the process, eliminate undesirable data and fine-tune parameters of the reduction. This step will result on the data validation or the appropriate parameters required to re-pipeline data in order to improve the quality of the final product.}
\end{itemize}





\section{Pipelines Approaches}
A pipeline is a collection of algorithms which are run in a particular order. The DCS will host pipelines coded in IDL, Python and other languages which are delivered by the instrument teams with a description specified as XML. With the appropriate pipeline specification, DCS can currently run pipelines in any language with no modification of the code as soon as the pipeline is delivered as an executable, likely the same that runs in SMO machines outside DCS. Because DCS does not have a knowledge of the details of the pipeline execution after it is called, we name this pipeline "blackbox". Level 2 blackboxes are applied based on the specifications of the AOR - which is detailed before the flight as part of the observation planning process – and the details of the process are recorded on the PP – which is created after pipelining  (Section 2). This approach is currently implemented in the DCS and embraces both automatic and user-initiated pipelining within the same framework. This answers the need for re-pipelining with the goal of improving the quality of the Level 2 data by fine tuning pipeline parameters after manual inspection or applying an improved version of the pipeline. Although, this approach represents an enormous cost saving on the implementation and maintenance of the pipelines it lacks the advanced functionalities that the DCS could offer including parallel execution of processes of a single pipeline, status report , and intermediate user intervention. 

We plan to complement the current functionality with another approach allowing human interaction. User interaction is required for step-by-step data processing and intermediate data analysis. These will be performed using a DCS graphical interface tool which runs user-interaction data process and analysis tools locally (outside DCS) after downloading updated algorithms from DCS. As a long term goal, DCS will integrate user-interaction pipelining within the same framework as automatic pipelining. For that purpose, pipelines will be delivered as a collection of functions (modules) performing a portion of the  pipeline and XML files describing them. The pipeline recipe (another XML file) will describe how modules are executed, the order of execution and how data is transfered between modules. Technically, the pipeline manager objects (pipe\_man) are in charge of executing specific modules (module-$>$process method) or the whole pipeline (pipe\_man-$>$run method). This new approach fits in the actual black box structure by calling run method as the pipeline executable. When implemented within DCS, pipe\_man will be able to process modules in parallel, control their execution, and allow user data analysis. In addition, pipe\_man will manage modules in different computer languages for the same pipeline thus reducing the number of algorithms in the system. Instrument teams will be encouraged to use existing algorithms when developing their pipelines, resulting in a common library of algorithms which will decrease the efforts of the instrument teams for developing pipelines and of the DCS team for maintaining and upgrading them. 

\section{Conclusion}
Combining automatic pipelining and user interaction of processing algorithms which are developed in various languages presents an important challenge to the SOFIA DCS --- especially when trying to minimize efforts required for long-term maintenance and upgrade of the code. We divide the problem in four distinct cases of interaction with the data. These scenarios can be developed independently but are based on a common architecture. The case of automatic pipelining, either at EoF or user-initiated, is already implemented and has been demonstrated with FLITECAM data. User-interactive pipelining is in its design phase but we have shown its feasibility using a prototype implemented in IDL. Flux calibration is not included in current DCS development plans due to resource/schedule constraints, but we provide the required tools for the user to ingest human validated data. 

\acknowledgements RYS is suported by USRA Contract to the Space Science Institute.  For more information about SOFIA visit http://www.sofia.usra.edu.


\end{document}